\documentclass[12pt]{article}

\usepackage[mathscr]{eucal}
\usepackage{amsfonts}
\usepackage{amssymb}
\usepackage{amsthm}

\newcommand{\be}{\begin{equation}}
\newcommand{\ee}{\end{equation}}
\newcommand{\bea}{\begin{eqnarray}}
\newcommand{\eea}{\end{eqnarray}}

\begin{document}

\begin{titlepage}
\vspace*{1.5cm}

\renewcommand{\thefootnote}{$\ddagger$}
\begin{center} {\LARGE\bf  Maxwell group and HS field theory }



\vspace{2cm}

{\large\bf Sergey Fedoruk},${}^{\dagger}$\footnote{\,\,On leave of absence from V.N.\,Karazin Kharkov National University, Ukraine}\qquad{\large\bf Jerzy
Lukierski} ${}^{\ast}$

\vspace{1.5cm}

${}^{\dagger}${\it Bogoliubov  Laboratory of Theoretical Physics, JINR,}\\ {\it Joliot-Curie 6, 141980 Dubna, Moscow region, Russia} \\ \vspace{0.1cm}

{\tt fedoruk@theor.jinr.ru}\\ \vspace{0.5cm}

${}^{\ast}${\it Institute for Theoretical Physics, University of Wroc{\l}aw,}\\ {\it pl. Maxa Borna 9, 50-204 Wroc{\l}aw, Poland} \\ \vspace{0.1cm}

{\tt lukier@ift.uni.wroc.pl}\\

\vspace{2cm}



\end{center} \vspace{0.2cm} 

\begin{abstract} \noindent
We consider the master fields for HS multiplets defined on 10-dimensional tensorial extension $\tilde{\cal M}$ of
$D=4$ space-time described as a coset $\tilde{\cal M}={\cal M}/Sl(2;\mathbb{C})$ of 16-parameter Maxwell group $\cal M$.
The tensorial coordinates provide a geometrization of the coupling to constant uniform EM fields.
We describe the spinorial model in extended space-time $\tilde{\cal M}$ and by its first quantization we obtain new
infinite HS-Maxwell multiplets with their massless components coupled to each other through constant EM background.
We conclude our report by observing that three-dimensional spinorial model
with a pair of spinors should provide after quantization $D=3$ massive HS-Maxwell multiplets.
\end{abstract}

\bigskip\bigskip \noindent PACS: 11.10.Ef, 11.30.Cp, 02.20.Sv

\smallskip \noindent Keywords: Maxwell symmetry, higher spins, tensorial space

\newpage

\end{titlepage}

\setcounter{footnote}{0}

\setcounter{equation}{0}
\section{Introduction}

\quad\, In order to introduce the field-theoretic description of infinite number of relativistic quantum fields with all spins
it is convenient to consider the enlargement of space-time.
There were proposed various extensions of Minkowski space-time with vectorial \cite{Frons1,Jerv}
and tensorial \cite{Frons2,BL,BLS,Vas,FedLuk} auxiliary coordinates, with master fields
describing the infinite-dimensional spin multiplets by Taylor expansion in additional variables.
In such a way one connects the master HS fields with enlarged $D=4$ Poincare algebra, with new generators describing the shifts in auxiliary variables.

In this talk we shall follow old derivation of HS field multiplets by quantization of spinorial superparticle model,
invariant under SUSY with six tensorial central charges \footnote{We point out that tensorial central charges are commuting
with all generators except the Lorentz generators.} \cite{BL,BLS,FedLuk}. Analogous tensorial charges occur e.g. in $D=11$
in M-algebra \cite{Sezg} which as it is postulated describes the algebraic structure of eleven-dimensional M-theory.
The master fields in $D=4$ are described in such approach by fields on extended space-time (${\rm x}_{\mu},{\rm z}_{\mu\nu}$),
where ${\rm z}_{\mu\nu}={\rm z}_{[\mu\nu]}$ are six auxiliary translations generated by tensorial central charges.
Further in order to describe integer and half-integer spins it is convenient to express the tensorial central
charges as bilinears in $D=4$ Weyl spinorial variables $\lambda_{\alpha}$, $\bar\lambda_{\dot\alpha}=(\overline{\lambda_{\alpha}})$,
$\alpha=1,2$
\footnote{We shall use $D=4$ two-spinor notation, i.e.
$P_{\alpha\dot\beta}=\sigma^{\mu}_{\alpha\dot\beta}P_{\mu}$, $Z_{\alpha\beta}=\sigma^{\mu\nu}_{\alpha\beta}Z_{\mu\nu}$, $\bar
Z_{\dot\alpha\dot\beta}=\tilde\sigma^{\mu\nu}_{\dot\alpha\dot\beta}Z_{\mu\nu}$, where
$(\sigma^\mu)_{\alpha\dot\alpha}=(1_2,\vec{\sigma})_{\alpha\dot\alpha}$,
$(\tilde\sigma^\mu)^{\dot\alpha\alpha}=\epsilon^{\alpha\beta}\epsilon^{\dot\alpha\dot\beta}(\sigma^\mu)_{\beta\dot\beta}=
(1_2,-\vec{\sigma})^{\dot\alpha\alpha}$, $\sigma^{\mu\nu}=i\,\sigma^{[\mu}\tilde\sigma^{\nu]}$, $\tilde\sigma^{\mu\nu}=i\,\tilde\sigma^{[\mu}\sigma^{\nu]}$
$\sigma^{\mu\nu}_{\alpha\beta}=\epsilon_{\beta\gamma}(\sigma^{\mu\nu})_{\alpha}{}^{\gamma}$,
$\tilde\sigma^{\mu\nu}_{\dot\alpha\dot\beta}=\epsilon_{\dot\beta\dot\gamma}(\tilde\sigma^{\mu\nu})^{\dot\gamma}{}_{\dot\alpha}$. We
use weight coefficient in (anti)symmetrization, i.e. $A_{(\alpha}B_{\beta)}=\frac12\,(A_{\alpha}B_{\beta}+A_{\beta}B_{\alpha})$,
$A_{[\alpha}B_{\beta]}=\frac12\,(A_{\alpha}B_{\beta}-A_{\beta}B_{\alpha})$. }
\begin{equation}\label{Z-spin-expr}
{\rm Z}_{\mu\nu} =  \lambda^{\alpha}(\sigma_{\mu\nu})_{\alpha}{}^{\beta}\lambda_{\beta}+
\bar\lambda_{\dot\alpha}(\tilde\sigma_{\mu\nu})^{\dot\alpha}{}_{\dot\beta}\bar\lambda^{\dot\beta}
\end{equation}
in analogy to the Penrose formula for massless fourmomenta
\begin{equation}\label{P-spin-expr}
{\rm P}_{\mu} = \lambda^{\alpha}(\sigma_{\mu})_{\alpha\dot\beta}\bar\lambda^{\dot\beta}\,.
\end{equation}
Finally we will arrive at master fields depending on spinorially extended space-time ($x_{\mu},\lambda_{\alpha},\bar\lambda_{\dot\alpha}$),
with their Taylor expansions describing HS fields with arbitrary spin.

The description of HS with tensorial coordinates has been further generalized by M.\,Vasiliev \cite{Vas,Vas1,Vas2} who followed
Fronsdal observation \cite{Frons2} that the infinite HS free multiplets can be described by single irrep of $D=4$
generalized conformal algebra $Sp(8)$, containing as its subalgebra the generalized Poincare algebra with tensorial central charges.
Subsequently the extended space-time method was applied to the description in AdS space-time and there were derived the multiplets of free HS
fields on AdS space \cite{PST1,PST2}. At present only two free HS field multiplets -- in Minkowski and AdS space-time --
are explicitly described, and both were derived by the method of quantization of spinorial model of Shirafuji type
\footnote{The formula (\ref{P-spin-expr}) was firstly inserted in massless spinorial Shirafuji model \cite{Shir}
which was the first one describing the link between the model
describing standard relativistic massless superparticle and the one describing free twistorial dynamics in
$D=4$ supertwistor space.

\quad The formula  (\ref{Z-spin-expr}) has been firstly employed in the superparticle model in \cite{BL}.}.

In our presentation we shall describe the spinorial model defined on other version of ten-dimensional extended
space-time ($x_{\mu},z_{\mu\nu}$), with coordinates $z_{\mu\nu}$ generated by the tensorial
central charges $Z_{\mu\nu}$ introduced in alternative way in the enlargement of Poincare algebra --
by replacing commuting four-momenta ${\rm P}_{\mu}$ by noncommutative ones $P_{\mu}$
\begin{equation}\label{PP-Z}
[{\rm P}_{\mu},{\rm P}_{\nu}] = 0 \qquad \Rightarrow\qquad [P_{\mu},P_{\nu}] = Z_{\mu\nu}
\end{equation}
where
\begin{equation}\label{PPZ}
[Z_{\mu\nu},Z_{\lambda\rho}] = 0\,, \qquad  [P_{\mu}, Z_{\lambda\rho}] = 0\,.
\end{equation}
Formulae  (\ref{PP-Z}), (\ref{PPZ}) describe the enlargement of $D=4$ Poincare algebra to $D=4$ Maxwell algebra
firstly proposed more than 40 years ago \cite{Bac,Schr,B}.
Using Maurer-Cartan (MC) forms for corresponding Maxwell group one can introduce \cite{FedLuk}
the Maxwell-invariant spinorial particle model and perform its first quantization.\footnote{
Different Maxwell-invariant particle model without using spinorial variables was made in \cite{FedLuk2}.
}
By analysis of the constraints in extended phase space we shall obtain in Schr\"{o}dinger realization
the master fields for Maxwell-HS free fields which are coupled to each other by constant uniform EM field.

The plan of our talk is the following. In Sect.\,2 we will recall the notion of Maxwell group,
Maxwell algebra and present the corresponding MC one-forms. Using this geometric framework we present in
Sect.\,3 the spinorial model \cite{FedLuk} and its first quantization with complete discussion of occurring
first and second class constraints. We will use the conversion method \cite{FS} which interprets canonical pair
of second class constraints as describing the system with gauge-fixed local gauge transformations
(one constraint from the canonical pair is considered as first class constraint and
generating the gauge transformations, second as introducing the
gauge-fixing condition). In such a way we get gauge-equivalent formulation of our model with nine first class constraints.
In Sect.\,4 we consider in detail the wave function of such a model satisfying on enlarged space-time nine wave equations.
Finally we express the solution of quantum-mechanical model in terms of local $D=4$ HS fields,
which will be named free Maxwell-HS fields. In such a way we obtain three infinite-dimensional coupled sets of
Maxwell-HS fields: one describing all HS bosonic fields ($s=0,1,2,\ldots$) and two infinite set
of chiral and antichiral fermionic HS fields with half-integer spins ($s=1/2,3/2,5/2,\ldots$).
In Sect.\,5 we will discuss the considered model in dual representation
as the one describing more explicitly Minkowski HS fields interacting with
constant EM field. Sect.\,6 is devoted to the outlook.
In particular we conjecture that suitable reduction $D=4$ $\rightarrow$ $D=3$ of considered in \cite{FedLuk} $D=4$
spinorial model can provide the infinite-dimensional $D=3$ Maxwell HS massive multiplets.

\setcounter{equation}{0}
\section{Maxwell algebra and covariant Maurer-Cartan one-forms}

\quad\,The Maxwell algebra \cite{Bac,Schr} is the semi-direct sum of Lorentz algebra with the generators
$M_{\alpha\beta}=M_{\beta\alpha}$, $\bar M_{\dot\alpha\dot\beta}=(M_{\alpha\beta})^+$ satisfying relations ${}^2$
\begin{equation}\label{Lor-alg}
\begin{array}{rrc}
&&[M_{\alpha\beta},M_{\gamma\delta}] = i\left( \epsilon_{\alpha\gamma}M_{\beta\delta}+
\epsilon_{\beta\delta}M_{\alpha\gamma}\right), \,\,\,\, [\bar M_{\dot\alpha\dot\beta},\bar M_{\dot\gamma\dot\delta}] = i\left(
\epsilon_{\dot\alpha\dot\gamma}\bar M_{\dot\beta\dot\delta}+ \epsilon_{\dot\beta\dot\delta}\bar M_{\dot\alpha\dot\gamma}\right),\\[7pt]
&&[M_{\alpha\beta},\bar M_{\dot\gamma\dot\delta}] = 0 \,,
\end{array}
\end{equation}
and the ten-dimension sector with generators of the Poincare translations
$P_{\alpha\dot\beta}=(P_{\beta\dot\alpha})^+$
and six tensorial generators $Z_{\alpha\beta}=Z_{\beta\alpha}$, $\bar
Z_{\dot\alpha\dot\beta}=(Z_{\alpha\beta})^+$.
The last ten generators transform under Lorentz algebra as it is indicated in the algebraic relations
\begin{equation}\label{M-alg1}
\begin{array}{rrc}
&&[M_{\alpha\beta},P_{\gamma\dot\gamma}] = -i\epsilon_{\gamma(\alpha}P_{\beta)\dot\gamma}\,,\qquad [\bar M_{\dot\alpha\dot\beta},P_{\gamma\dot\gamma}] =
-i\epsilon_{\dot\gamma(\dot\alpha}P_{\gamma\dot\beta)}\,, \\[7pt]
&&[M_{\alpha\beta},Z_{\gamma\delta}] = i\left( \epsilon_{\alpha\gamma}Z_{\beta\delta}+
\epsilon_{\beta\delta}Z_{\alpha\gamma}\right), \quad [\bar M_{\dot\alpha\dot\beta},\bar Z_{\dot\gamma\dot\delta}] = i\left(
\epsilon_{\dot\alpha\dot\gamma}\bar Z_{\dot\beta\dot\delta}+ \epsilon_{\dot\beta\dot\delta}\bar Z_{\dot\alpha\dot\gamma}\right), \\[7pt]
&&[M_{\alpha\beta},\bar Z_{\dot\gamma\dot\delta}] = 0,\quad [\bar M_{\dot\alpha\dot\beta},Z_{\gamma\delta}] = 0,
\end{array}
\end{equation}
Differently than in the Poincare algebra with commuting translation operators,
the commutators of the quantities $P_{\alpha\dot\beta}$ yield new tensorial generators
(see also (\ref{PP-Z}))
\begin{equation}\label{M-alg}
[P_{\alpha\dot\alpha},P_{\beta\dot\beta}] = 2 i\,e \left(\epsilon_{\dot\alpha\dot\beta}Z_{\alpha\beta}+
\epsilon_{\alpha\beta}\bar Z_{\dot\alpha\dot\beta}\right),
\end{equation}
where $e$ is a constant interpreted further as describing EM coupling.
Six tensorial charges $Z_{\alpha\beta}$, $\bar Z_{\dot\alpha\dot\beta}$ commute with each other and with the
Poincare translation generators, i.e. the following commutators (see also (\ref{PPZ}))
\begin{equation}\label{M-algZ}
[Z_{\alpha\beta},Z_{\gamma\delta}] = [Z_{\alpha\beta},\bar Z_{\dot\alpha\dot\beta}]=
[Z_{\alpha\beta},P_{\gamma\dot\gamma}]=0
\end{equation}
are valid.

The $D=4$ Maxwell algebra define the Maxwell group $\cal{M}$, in standard way by exponential representation.
We define the $D=4$ proper Maxwell group ${\cal{M}}_0$, determining Maxwell tensorial space, as ten-dimensional coset
\begin{equation}\label{M-coset}
{\cal{M}}_0=\frac{\cal{M}}{O(3,1)}=e^{i(z^{\alpha\beta}Z_{\alpha\beta}+\bar z^{\dot\alpha\dot\beta}\bar Z_{\dot\alpha\dot\beta})}
e^{ix^{\alpha\dot\beta}P_{\alpha\dot\beta}}
\end{equation}
with generators $P_{\alpha\dot\beta}$, $Z_{\alpha\beta}$, $\bar Z_{\dot\alpha\dot\beta}$.
The coset coordinates have the following transformations under the space-time translations (parameters $a^{\alpha\dot\beta}$) and tensorial Maxwell
shifts (parameters $b^{\alpha\beta}$, $\bar b^{\dot\alpha\dot\beta}$)
\begin{equation}\label{M-trans}
\begin{array}{ccc} &&\delta x^{\alpha\dot\beta}= a^{\alpha\dot\beta}\,,\\[5pt] &&\delta
z^{\alpha\beta}=   b^{\alpha\beta} + x^{(\alpha\dot\gamma}a^{\beta)}_{\dot\gamma}\,,\qquad \delta\bar z^{\dot\alpha\dot\beta}= \bar
b^{\dot\alpha\dot\beta}+ x^{\gamma(\dot\alpha}a^{\dot\beta)}_{\gamma} \,.
\end{array}
\end{equation}
The Lorentz transformations
with the parameters $\ell^{\alpha\beta}$, $\bar\ell^{\dot\alpha\dot\beta}$ look as follows
\begin{equation}\label{M-trans-L}
\delta x^{\alpha\dot\beta}=
\ell^{\alpha\gamma}x_{\gamma}^{\dot\beta}+\bar\ell^{\dot\alpha\dot\gamma}x_{\dot\gamma}^{\beta}\,,\qquad \delta z^{\alpha\beta}=
2\ell^{\alpha\gamma}z_{\gamma}^{\beta} \,,\qquad \delta\bar z^{\dot\alpha\dot\beta}= 2\bar\ell^{\dot\alpha\dot\beta}\bar z_{\dot\gamma}^{\dot\beta}\,.
\end{equation}

Using the parametrization (\ref{M-coset}) and the algebraic relations (\ref{M-alg}), (\ref{M-algZ}) we can define the Maurer-Cartan (MC) one-forms
\begin{equation}\label{om-def}
{\cal M}_0^{-1}d {\cal M}_0=i\Big(\omega^{\alpha\dot\beta}P_{\alpha\dot\beta}+\omega^{\alpha\beta}Z_{\alpha\beta} + \bar\omega^{\dot\alpha\dot\beta}\bar Z
_{\dot\alpha\dot\beta} \Big)\,.
\end{equation}
Explicit formulae for MC one-forms defined by (\ref{om-def}) are
\begin{equation}\label{M-om}
\begin{array}{ccc}
&&\omega^{\alpha\dot\beta}= d x^{\alpha\dot\beta}\,,\\[5pt] &&\omega^{\alpha\beta}=  d z^{\alpha\beta} + e\,x^{(\alpha\dot\gamma}d
x^{\beta)}_{\dot\gamma}\,,\qquad \bar\omega^{\dot\alpha\dot\beta}= d\bar z^{\dot\alpha\dot\beta}+ e\,x^{\gamma(\dot\alpha}d x^{\dot\beta)}_{\gamma}\,.
\end{array}
\end{equation}
Corresponding covariant derivatives have the form
\begin{equation}\label{M-der}
\begin{array}{ccc}
&& D_{\alpha\dot\beta}= -i\left(\displaystyle\frac{\partial}{\partial
x^{\alpha\dot\beta}} \,\,+ e\,x^{\gamma}_{\dot\beta}\,\displaystyle\frac{\partial}{\partial z^{\alpha\gamma}}\,\,+
e\,x^{\dot\gamma}_{\alpha}\,\displaystyle\frac{\partial}{\partial \bar z^{\dot\beta\dot\gamma}}\right),\\[9pt] && D_{\alpha\beta}=
-i\,\displaystyle\frac{\partial}{\partial z^{\alpha\beta}}\,,\qquad \bar D_{\dot\alpha\dot\beta}= -i\,\displaystyle\frac{\partial}{\partial \bar
z^{\dot\alpha\dot\beta}}\,\,.
\end{array}
\end{equation}
Since the space-time and tensorial translations are the shifts of the group parameters in Maxwell tensorial space, they do not change
the MC one-forms (\ref{M-om}) and the form of operators (\ref{M-der}).
We add that the Lorentz symmetry acts on MC one-forms and covariant derivatives in standard way, by the linear transformations.

\setcounter{equation}{0}
\section{Particle action, constraints and the Casimirs}

\quad\, We will consider the model of massless HS particle which propagates in the Maxwell tensorial space $X
=(x^{\alpha\dot\alpha},z^{\alpha\beta}, \bar z^{\dot\alpha\dot\beta} )$ enlarged by the pair of spinorial variables.
Our model is described by the following Maxwell-invariant particle action
\begin{equation}\label{Maxwell-actT}
S=S_X+S_\lambda\,.
\end{equation}

First term in (\ref{Maxwell-actT})
\begin{equation}\label{Maxwell-act}
S_X=\int \,
\left(\lambda_{\alpha}\bar\lambda_{\dot\beta}\,\omega^{\alpha\dot\beta}+ a\lambda_{\alpha}\lambda_{\beta}\,\omega^{\alpha\beta} +
\bar a\bar\lambda_{\dot\alpha}\bar\lambda_{\dot\beta}\,\bar\omega^{\dot\alpha\dot\beta} \right)\,.
\end{equation}
is the generalization of the $D=4$ spinorial particle model defined on flat tensorial space-time in
\cite{BL,BLS,Vas} (it is a tensorial generalization of Shirafuji model \cite{Shir}).
The components of the {\it commuting} Weyl spinor $\lambda_{\alpha}$, $\bar\lambda_{\dot\alpha}=(\overline{\lambda_{\alpha}})$
can be further considered as parts of the $D=4$ twistor, and the model can be rewritten also as free $D=4$ twistor
particle model \cite{BL,BLS}.

In the action (\ref{Maxwell-act}) the parameter $a$ is complex.
Because $e$ in (\ref{Maxwell-act}) is dimensionless, we should choose
the tensorial coordinates $(z^{\alpha\beta},\bar z^{\dot\alpha\dot\beta})$  having mass dimensionality equal to $-2$,
$[z^{\alpha\beta}]=[\bar z^{\dot\alpha\dot\beta}]=-2$. Further from (\ref{Maxwell-act}) follows that
$[\lambda^{\alpha}]=[\bar\lambda^{\dot\alpha}]=\frac12$, and one can deduce that the complex parameter $a$ is mass-like, $[a]=1$.
This mass-like parameter can be chosen  real, $a=\bar a=m$, if we take into account $U(1)$ phase transformations of the spinors
$\lambda_{\alpha}=e^{i\varphi}\lambda_{\alpha}$, $\bar\lambda_{\dot\alpha}=e^{-i\varphi}\bar\lambda_{\dot\alpha}$.

Second term in (\ref{Maxwell-actT}) has the form
\begin{equation}\label{Maxwell-act1}
S_\lambda=\int \,
\Big(\lambda_{\alpha}d y^{\alpha} + {\bar\lambda}_{\dot\alpha}d{\bar y}^{\dot\alpha} \Big)\,.
\end{equation}
Such term defines the variables $(y^{\alpha},{\bar y}^{\dot\alpha})$, ${\bar y}^{\dot\alpha}=(\overline{y^{\alpha}})$, and
$(\lambda_{\alpha}$, $\bar\lambda_{\dot\alpha})$ as canonically conjugated pairs.
One can show that without second term (\ref{Maxwell-act1}),
the action (\ref{Maxwell-act}) possesses
a complicated structure of the constrains: in such a case follow
the second class constraints $y^{\alpha}\approx0$, $\bar y^{\dot\alpha}\approx0$
which make the quantization quite difficult \cite{BL,BLS}.
Adding the term (\ref{Maxwell-act1}) removes these complications.

Thus, the full Lagrangian of the model considered here is defined by \cite{FedLuk}
\begin{equation}\label{Maxwell-Lagr}
L=\lambda_{\alpha}\bar\lambda_{\dot\beta}\,\dot
x^{\alpha\dot\beta}+ m\lambda_{\alpha}\lambda_{\beta}\,\Big(\dot z^{\alpha\beta} + e\,x^{\alpha\dot\gamma}\dot x^{\beta}_{\dot\gamma}\Big) +
m\bar\lambda_{\dot\alpha}\bar\lambda_{\dot\beta}\,\Big( \dot{\bar z}^{\dot\alpha\dot\beta}+ e\,x^{\gamma\dot\alpha}\dot x^{\dot\beta}_{\gamma} \Big)+
\lambda_{\alpha}\dot y^{\alpha} + {\bar\lambda}_{\dot\alpha}\dot{\bar y}^{\dot\alpha}
\end{equation}
and describes the particle motion in generalized
coordinate space $(x^{\alpha\dot\beta},z^{\alpha\beta},{\bar z}^{\dot\alpha\dot\beta}, y^{\alpha},\bar y^{\dot\alpha})$, with generalized momenta
$(p_{\alpha\dot\beta},f_{\alpha\beta},{\bar f}_{\dot\alpha\dot\beta}, \lambda_{\alpha},\bar\lambda_{\dot\alpha})$
having the following nonvanishing Poisson brackets~(PB)
\begin{equation}\label{PB-c1}
\{x^{\alpha\dot\alpha},p_{\beta\dot\beta}\}_{{}_P}=\delta^{\alpha}_{\beta}\delta^{\dot\alpha}_{\dot\beta},\qquad
\{z^{\alpha\beta},f_{\gamma\delta}\}_{{}_P}=\delta^{(\alpha}_{\gamma}\delta^{\beta)}_{\delta},\qquad \{\bar z^{\dot\alpha\dot\beta},\bar
f_{\dot\gamma\dot\delta}\}_{{}_P}=\delta^{(\dot\alpha}_{\dot\gamma}\delta^{\dot\beta)}_{\dot\delta},
\end{equation}
\begin{equation}\label{PB-c2}
\{y^\alpha, \lambda_{\beta}\}_{{}_P}=\delta_{\beta}^\alpha, \qquad \{\bar y^{\dot\alpha},\bar\lambda_{\dot\beta} \}_{{}_P}=\delta_{\dot\beta}^{\dot\alpha}\,.
\end{equation}

The canonical definitions of vectorial and tensorial
momenta $(p_{\alpha\dot\beta},f_{\alpha\beta},{\bar f}_{\dot\alpha\dot\beta})$ provide from the action density (\ref{Maxwell-Lagr}) the constraints
\begin{eqnarray}\label{costr-x}
T_{\alpha\dot\beta}&\equiv&p_{\alpha\dot\beta}+
em\,\lambda_{\alpha}\lambda_{\gamma}x^\gamma_{\dot\beta}+
em\,\bar\lambda_{\dot\beta}\bar\lambda_{\dot\gamma}x^{\dot\gamma}_{\alpha}-\lambda_{\alpha}\bar\lambda_{\dot\beta}\approx 0\, ,\\[5pt]
T_{\alpha\beta}&\equiv&f_{\alpha\beta}-m\lambda_{\alpha}\lambda_{\beta}\approx 0\,,\label{costr-y}\\[5pt] \bar T_{\dot\alpha\dot\beta}&\equiv&\bar
f_{\dot\alpha\dot\beta}-m\bar\lambda_{\dot\alpha}\bar\lambda_{\dot\beta}\approx 0\,.\label{costr-by}
\end{eqnarray}

In first order formulation we can introduce in spinorial sector of the model (\ref{Maxwell-Lagr}) ``momenta for momenta'' and consider the constraints
$p_\lambda^{\alpha}-y^{\alpha}\approx0$, $\bar p_\lambda^{\dot\alpha} -\bar
y^{\dot\alpha}\approx0$ and $p_{y\,\alpha}\approx0$, $\bar p_{y\,\dot\alpha}\approx0$.
Subsequent introduction of Dirac brackets eliminates however the variables
$p_\lambda$, $\bar p_\lambda$, $p_{y}$, $\bar p_{y}$ from the phase space and we get only PB (\ref{PB-c2}).

Inserting (\ref{costr-y}), (\ref{costr-by}) in (\ref{costr-x}) we obtain the following
representation of the constraints (\ref{costr-x})-(\ref{costr-by})
\begin{equation}\label{costr-n}
T_{\alpha\dot\beta}= D_{\alpha\dot\beta}-\lambda_{\alpha}\bar\lambda_{\dot\beta}\approx 0\,,\qquad
T_{\alpha\beta}= D_{\alpha\beta}-\lambda_{\alpha}\lambda_{\beta}\approx 0\,,\qquad
T_{\dot\alpha\dot\beta}= D_{\dot\alpha\dot\beta}-\bar\lambda_{\dot\alpha}\bar\lambda_{\dot\beta}\approx 0\,,
\end{equation}
where
\begin{equation}\label{D-cov-cl}
D_{\alpha\dot\beta}=p_{\alpha\dot\beta} +e\,x^\gamma_{\dot\beta}f_{\alpha\gamma}+ e\,x^{\dot\gamma}_{\alpha}\bar
f_{\dot\beta\dot\gamma}\,,\qquad D_{\alpha\beta}=f_{\alpha\beta} \,,\qquad
D_{\dot\alpha\dot\beta}=\bar f_{\dot\alpha\dot\beta}
\end{equation}
are the classical counterparts of the Maxwell-covariant derivatives (\ref{M-der}).
We see that the constraints (\ref{costr-n}) are the Maxwell-covariant generalization
of the constraints leading to so-called unfolded equations for HS fields \cite{BLS,Vas}.

We stress that present model has important difference in comparison with the one describing the HS particle of \cite{BLS,Vas},
because the vectorial constraints do not commute (do have nonvanishing PB).
There are the following nonvanishing Poisson brackets of the vectorial constraints (\ref{costr-x})
\begin{equation}\label{costr-PB}
\left\{T_{\alpha\dot\alpha}, T_{\beta\dot\beta}\right\}_{{}_P}= 2 e\left(\epsilon_{\dot\alpha\dot\beta} f_{\alpha\beta}+ \epsilon_{\alpha\beta}\bar
f_{\dot\alpha\dot\beta}\right) \approx 2e\, m \left(\epsilon_{\dot\alpha\dot\beta} \lambda_{\alpha}\lambda_{\beta}+
\epsilon_{\alpha\beta}\bar\lambda_{\dot\alpha}\bar\lambda_{\dot\beta}\right).
\end{equation}
Other tensorial constraints (\ref{costr-y}), (\ref{costr-by}) are the same as in \cite{BLS,Vas} and
commute with all the constraints (\ref{costr-x})-(\ref{costr-by}).
Thus, the tensorial constraints (\ref{costr-y}), (\ref{costr-by}) are first class
whereas the vectorial constraints (\ref{costr-x}) are the superposition of two first class and two second class constraints.
In limit $e\to 0$ the Lagrangian (\ref{Maxwell-Lagr}) becomes the HS particle Lagrangian from
\cite{BLS,Vas} where the covariant derivatives (\ref{D-cov-cl}) form Abelian algebra and all constraints (\ref{costr-x})-(\ref{costr-by}) are first class.

To perform quantization of our model it is important to project out the first and second class constraints present in (\ref{costr-x}).
If we wish to preserve the Lorentz covariance this separation requires the use of additional spinorial variables.
In order to have a basis in two-dimensional spinor space, we introduce second spinor, $u_{\alpha}$, as firstly proposed in \cite{BLS}.
This auxiliary spinor satisfies the normalization condition
\begin{equation}\label{norm-u}
\lambda^\alpha u_{\alpha}=1\,.
\end{equation}
The nonvanishing PBs of $u_{\alpha}$
\begin{equation}\label{PB-u}
\{u_{\alpha},y^\beta \}_{{}_P}=u_{\alpha}u^\beta
\end{equation}
preserve the normalization  (\ref{norm-u}).

Using this spinorial basis $(\lambda_{\alpha},u_{\alpha})$ we can introduce in Lorentz-covariant way the following projections
\begin{equation}\label{costr-pr}
T_{\lambda\bar\lambda}\equiv \lambda^\alpha
T_{\alpha\dot\alpha}\bar\lambda^{\dot\alpha}\,, \qquad T_{\lambda\bar u}\equiv \lambda^\alpha T_{\alpha\dot\alpha}\bar u^{\dot\alpha}\,, \qquad
T_{u\bar\lambda}\equiv u^\alpha T_{\alpha\dot\alpha}\bar\lambda^{\dot\alpha}\,, \qquad T_{u\bar u}\equiv u^\alpha T_{\alpha\dot\alpha}\bar
u^{\dot\alpha}\,.
\end{equation}
Their unique nonvanishing PB is the following
\begin{equation}\label{PB-pr}
\{T_{\lambda\bar u}+T_{u\bar\lambda}, T_{u\bar u}\}_{{}_P}=4e\,m
\end{equation}
and one can conclude that the constraints
\begin{equation}\label{1-cl}
T_{\lambda\bar u}-T_{u\bar\lambda}\,,\qquad
T_{\lambda\bar\lambda}
\end{equation}
are first class, whereas
\begin{equation}\label{2-cl}
T_{\lambda\bar u}+T_{u\bar\lambda}\,,\qquad
T_{u\bar u}
\end{equation}
are second class.

Introduction of the Dirac brackets for the second class constraints leads to complicated structure of
the quantum-mechanical algebra. As a way out we use the conversion method \cite{FS} in which the canonical pair
of second class constraints can be considered as system where one second class constraint is interpreted as gauge-fixing condition
for the gauge transformations which are generated by the other constraint. In the following we will consider the constraint $T_{u\bar u}\approx0$
as gauge fixing condition, and the constraint $T_{\lambda\bar u}+T_{u\bar\lambda}\approx0$ as generating new gauge degree of freedom.
Finally we will consider the classical gauge-equivalent system which is described by the constraints
\begin{equation}\label{costr-1tr}
T_{\lambda\bar u}\approx0\,,\qquad T_{u\bar\lambda}\approx0\,, \qquad
T_{\lambda\bar\lambda}\approx0
\end{equation}
replacing the vectorial constraints (\ref{costr-x}).

The constraints (\ref{costr-1tr}) are in fact linear combinations of
the projections $T_{\alpha\dot\beta}\bar\lambda^{\dot\beta}$, $\lambda^{\beta}T_{\beta\dot\alpha}$
of the constraints  (\ref{costr-x}) on the Weyl spinor components
$\lambda^\alpha$, $\bar\lambda^{\dot\alpha}$.
So, we can avoid using the auxiliary spinor $u_{\alpha}$ in definition of the constraints and as the
result, we describe our model by
the following set of first class constraints
\begin{eqnarray}\label{costr-x1}
S_{\alpha}&\equiv&T_{\alpha\dot\beta}\bar\lambda^{\dot\beta}=\left( p_{\alpha\dot\beta}+ e\,f_{\alpha\gamma}x^{\gamma}_{\dot\beta}
\right)\bar\lambda^{\dot\beta} \approx D_{\alpha\dot\beta}\bar\lambda^{\dot\beta}\approx 0\, ,\\[5pt] \label{costr-x1a} \bar
S_{\dot\alpha}&\equiv&\lambda^{\beta}T_{\beta\dot\alpha}=\lambda^{\beta}\left( p_{\beta\dot\alpha}+ e\,\bar f_{\dot\alpha\dot\gamma}
x_{\beta}^{\dot\gamma}\right) \approx \lambda^{\beta} D_{\beta\dot\alpha}\approx 0\, ,\\[5pt]
T_{\alpha\beta}&\equiv&f_{\alpha\beta}-m\lambda_{\alpha}\lambda_{\beta}\approx 0\,,\label{costr-y1}\\[5pt] \bar T_{\dot\alpha\dot\beta}&\equiv&\bar
f_{\dot\alpha\dot\beta}-m\bar\lambda_{\dot\alpha}\bar\lambda_{\dot\beta}\approx 0\,.\label{costr-by1}
\end{eqnarray}
We observe that the four constraints (\ref{costr-x1}), (\ref{costr-x1a}) are not independent, because they satisfy the relation
\begin{equation}\label{dep}
\lambda^{\alpha}S_{\alpha}=\bar\lambda^{\dot\alpha}\bar S_{\dot\alpha}\,,
\end{equation}
i.e. we get only three independent first class constraints.
It appears that the condition (\ref{dep}) does not enter into the derivation of the spectrum of our model.

In the transition to this new system of the constraints, we should be careful not to loose any of the constraints.
In particular, performing the projections (\ref{costr-x1}), (\ref{costr-x1a}) we are omitting the contribution to the
vectorial constraint (\ref{costr-x}) which does not depend on spinor variables.
Such contribution is described by the following new constraint
\begin{equation}\label{constr-x3}
T\equiv T_{\alpha\dot\beta}T^{\alpha\dot\beta}\approx 0\,.
\end{equation}
This quadratic constraint is of first class. Indeed, the constraint (\ref{constr-x3}) can be represented by the formula
$T=T_{\lambda\bar\lambda}T_{u\bar u}-T_{\lambda\bar
u}T_{u\bar\lambda}\approx 0$ and therefore due to (\ref{costr-1tr}) it is first class.
Thus, we should add the constraint (\ref{constr-x3}) to the first class constraints (\ref{costr-x1})-(\ref{costr-by1}).
After quantization this constraint will provide the Maxwell extension of the
Klein-Gordon (KG) equation.

One can find a physical interpretation of the system of the first class constraints
(\ref{costr-x1})-(\ref{costr-by1}), (\ref{constr-x3}). For that purpose we will look for
the values of the Casimir operators for the symmetry algebra in our model,
i.e. the Casimirs of the the Maxwell algebra \cite{Schr,Sor,BGKL}
\begin{eqnarray}\label{Max-Cas1} C^{Max}_1 &=& P_{\alpha\dot\beta}P^{\alpha\dot\beta} +4e\Big(M_{\alpha\beta}
Z^{\alpha\beta} +\bar M_{\dot\alpha\dot\beta}\bar Z^{\dot\alpha\dot\beta} \Big)\,, \\[3pt] \label{Max-Cas23} C^{Max}_2 &=& Z_{\alpha\beta} Z^{\alpha\beta}
\,,\qquad\qquad C^{Max}_3 =\bar Z_{\dot\alpha\dot\beta}\bar Z^{\dot\alpha\dot\beta}\,, \\[3pt] \label{Max-Cas4} C^{Max}_4 &=& 2Z^{\alpha\beta}\bar
Z^{\dot\alpha\dot\beta}P_{\alpha\dot\alpha}P_{\beta\dot\beta}  - {\textstyle\frac12}\left(Z^{\gamma\delta}Z_{\gamma\delta}+ \bar
Z^{\dot\gamma\dot\delta}\bar Z_{\dot\gamma\dot\delta}\right) P^{\alpha\dot\alpha}P_{\alpha\dot\alpha}\\ &&\qquad\qquad\qquad\quad\,\,\, +\,
2e\left(Z^{\gamma\delta}Z_{\gamma\delta}- \bar Z^{\dot\gamma\dot\delta}\bar Z_{\dot\gamma\dot\delta}\right) \left(Z^{\alpha\beta}M_{\alpha\beta}- \bar
Z^{\dot\alpha\dot\beta}\bar M_{\dot\alpha\dot\beta}\right). \nonumber
\end{eqnarray}

Using the transformations (\ref{M-trans}), (\ref{M-trans-L}) and the following transformations of spinors
\begin{equation}\label{M-trans-L1}
\delta y^{\alpha}=  \ell^{\alpha\beta}y_{\beta} \,,\quad \delta\bar
y^{\dot\alpha}= \bar\ell^{\dot\alpha\dot\beta}\bar y_{\dot\beta}\,,\qquad \delta \lambda_{\alpha}=  -\ell_{\alpha\beta}\lambda^{\beta} \,,\quad \delta\bar
\lambda_{\dot\alpha}= -\bar\ell_{\dot\alpha\dot\beta}\bar\lambda^{\dot\beta}
\end{equation}
we obtain from the action (\ref{Maxwell-Lagr}) the Noether charges in our model
\begin{equation}\label{M-gen-cl}
\begin{array}{rcl} P_{\alpha\dot\beta}&=&-p_{\alpha\dot\beta}+e\,x_{\alpha}^{\dot\gamma}\bar
f_{\dot\gamma\dot\beta}+e\,f_{\alpha\gamma} x_{\dot\beta}^{\gamma}\,,\\[5pt] Z_{\alpha\beta}&=&-f_{\alpha\beta}\,,\qquad \qquad \bar
Z_{\dot\alpha\dot\beta}\,\,\,=\,\,\,-\bar f_{\dot\alpha\dot\beta}\,,\\[5pt] M_{\alpha\beta}&=&x_{(\alpha}^{\dot\gamma}p_{\beta)\dot\gamma} +
2z_{(\alpha}^{\gamma}f_{\beta)\gamma} +y_{(\alpha}\lambda_{\beta)}\,,\\[5pt] \bar M_{\dot\alpha\dot\beta}&=&x_{(\dot\alpha}^{\gamma}p_{\gamma\dot\beta)} +
2\bar z_{(\dot\alpha}^{\dot\gamma}\bar f_{\dot\beta)\dot\gamma} +\bar y_{(\dot\alpha}\bar \lambda_{\dot\beta)}\,,
\end{array}
\end{equation}
which represent the dynamical phase space realization of Maxwell algebra generators.
If we insert the expressions (\ref{M-gen-cl}) in the Casimirs (\ref{Max-Cas1}), (\ref{Max-Cas23}), (\ref{Max-Cas4})
and take into account the first class constraints (\ref{costr-x1})-(\ref{costr-y1}), (\ref{constr-x3}) we find  that
\begin{equation}\label{Cas-fix}
C^{Max}_1 \approx0\,,\qquad C^{Max}_2
\approx0 \,,\qquad C^{Max}_3 \approx0\,,\qquad C^{Max}_4
\approx0\,.
\end{equation}
Thus, our Maxwell-Shirafuji model, defined by the first class constrains (\ref{costr-x1})-(\ref{costr-y1}), (\ref{constr-x3}),
describes the Maxwell-HS particle multiplet with all vanishing eigenvalues of Casimirs (\ref{Max-Cas1})-(\ref{Max-Cas4}),
similarly as massless scalar relativistic particle model is characterized by vanishing Casimirs of Poincare algebra.

\setcounter{equation}{0} \section{First quantization of the particle model and  interacting HS fields}

\quad\, Phase space coordinates after quantization become the operators, and for simplicity we will denote them further by the same letter (without hats).
The Poisson brackets algebra (\ref{PB-c1}), (\ref{PB-c2}) generates
the following quantum-mechanical algebra
\begin{equation}\label{CR-c1}
[x^{\alpha\dot\alpha},p_{\beta\dot\beta}]=i\delta^{\alpha}_{\beta}\delta^{\dot\alpha}_{\dot\beta},\qquad
[z^{\alpha\beta},f_{\gamma\delta}]=i\delta^{(\alpha}_{\gamma}\delta^{\beta)}_{\delta},\qquad [\bar z^{\dot\alpha\dot\beta},\bar
f_{\dot\gamma\dot\delta}]=i\delta^{(\dot\alpha}_{\dot\gamma}\delta^{\dot\beta)}_{\dot\delta},
\end{equation}
\begin{equation}\label{CR-c2}
[y^\alpha, \lambda_{\beta}]=i\delta_{\beta}^\alpha, \qquad [\bar y^{\dot\alpha},\bar\lambda_{\dot\beta} ]=i\delta_{\dot\beta}^{\dot\alpha}\,.
\end{equation}

Further we consider Schr\"{o}dinger-type representation in which the operators
$x^{\alpha\dot\beta}$, $z^{\alpha\beta}$, $\bar z^{\dot\alpha\dot\beta}$, $y^\alpha,\bar y^{\dot\alpha}$
are realized as multiplications by $c$-numbers whereas
the operators of quantized momenta are represented as partial derivatives
\begin{equation}\label{op-real1-a}
p_{\alpha\dot\beta}=-i\frac{\partial\,\,}{\partial x^{\alpha\dot\beta}}\equiv
-i\partial_{\alpha\dot\beta}\,,\qquad
\lambda_{\alpha}=-i\frac{\partial\,\,}{\partial y^{\alpha}}\equiv  -i\partial_{\alpha}\,,\qquad
\bar\lambda_{\dot\alpha}=-i\frac{\partial\,\,}{\partial \bar y^{\dot\alpha}}\equiv  -i\bar\partial_{\dot\alpha}\,,
\end{equation}
\begin{equation}\label{op-real2-a}
f_{\alpha\beta}=-i\frac{\partial\,\,}{\partial z^{\alpha\beta}}\equiv -i\partial_{\alpha\beta}\,,\qquad \bar
f_{\dot\alpha\dot\beta}=-i\frac{\partial\,\,}{\partial \bar z^{\dot\alpha\dot\beta}}\equiv -i\bar\partial_{\dot\alpha\dot\beta}
\,.
\end{equation}
The physical spectrum of the wave function
\begin{equation}\label{WF-a}
\Phi =\Phi(x^{\alpha\dot\beta},z^{\alpha\beta}, \bar z^{\dot\alpha\dot\beta}, y^\alpha,\bar y^{\dot\alpha})
\end{equation}
is defined by the quantum counterpart of first class constraints  (\ref{costr-x1})-(\ref{costr-y1}):
\begin{eqnarray}\label{q-costr-x1-a}
i D_{\alpha\dot\beta}\bar\partial^{\dot\beta}\,\Phi&=&\left(\partial_{\alpha\dot\beta}+
e\,\partial_{\alpha\gamma}\,x_{\dot\beta}^{\gamma}\right)\bar\partial^{\dot\beta}\,\Phi= 0\, ,\\[5pt] \label{q-costr-z1-a}
i D_{\beta\dot\alpha}\partial^{\beta}\,\Phi&=&\left(\partial_{\beta\dot\alpha}+ e\,\bar \partial_{\dot\alpha\dot\gamma} \,x_{\beta}^{\dot\gamma}
\right)\partial^{\beta}\, \Phi= 0\, ,
\end{eqnarray}
\begin{equation}\label{q-costr-y1-a}
\Big(\partial_{\alpha\beta}+ i m\partial_{\alpha}
\partial_{\beta}\Big)\Phi=0\,,\qquad\qquad \Big(\bar \partial_{\dot\alpha\dot\beta}+ im \bar\partial_{\dot\alpha}\bar\partial_{\dot\beta}\Big)\Phi=0\,,
\end{equation}
where $D_{\alpha\dot\beta}$ is the Maxwell-covariant derivative (see (\ref{D-cov-cl}), (\ref{M-der})).

The solutions of eqs.\,(\ref{q-costr-y1-a}) are described by compact formula
\begin{equation}\label{WF-1-comp1}
\Phi(x,z, \bar z, y,\bar
y)=e^{-im\left(z^{\alpha\beta}\partial_{\alpha} \partial_{\beta}+\bar z^{\dot\alpha\dot\beta}\bar\partial_{\dot\alpha}\bar\partial_{\dot\beta}\right)}
\Phi_{0}(x, y,\bar y)\,.
\end{equation}
From expression (\ref{WF-1-comp1}) follows that the tensorial coordinates $z^{\mu\nu}=(z^{\alpha\beta},\bar
z^{\dot\alpha\dot\beta})$ are the auxiliary gauge degrees of freedom and the gauge-independent degrees of freedom are present in
the HS master field $\Phi_{0}(x, y,\bar y)$. Residual equations (\ref{q-costr-x1-a}), (\ref{q-costr-z1-a}) for unconstrained field $\Phi_{0}(x, y,\bar y)$
take the form
\begin{eqnarray}\label{q-costr-x1-c}
\left[\partial_{\alpha\dot\beta}\bar\partial^{\dot\beta}+iem\,
(\partial^{\beta}x_{\beta\dot\beta}\bar\partial^{\dot\beta})\partial_{\alpha}\right]\Phi_0&=& 0\, ,\\[5pt] \label{q-costr-z1-c}
\left[\partial^{\beta}\partial_{\beta\dot\alpha}+iem\,
(\partial^{\beta}x_{\beta\dot\beta}\bar\partial^{\dot\beta})\bar \partial_{\dot\alpha}
\right] \Phi_0&=& 0\, .
\end{eqnarray}

Spinor variables $ y^{\alpha}$, $\bar y^{\dot\alpha}$ are besides space-time coordinates, the additional spinorial variables.
We consider the following Taylor expansion of HS master field with respect to the additional spinor variables
\begin{equation}\label{WF-0-a}
\Phi_{0}(x, y,\bar y)=
\sum_{k,n=0}^{\infty} \frac{1}{k!n!}\, y^{\alpha_1}...y^{\alpha_k}\, \bar y^{\dot\beta_1}...\bar y^{\dot\beta_n}
\,\phi^{(k,n)}_{\alpha_1...\alpha_k\,\dot\beta_1...\dot\beta_n}(x)\,,
\end{equation}
where Maxwell-HS component fields
$\phi^{(k,n)}_{\alpha_1...\alpha_k\,\dot\beta_1...\dot\beta_n}(x)$ are the usual $D=4$ space-time spin-tensor fields.

Let us find now the solution of the equations (\ref{q-costr-x1-c}), (\ref{q-costr-z1-c}).

In the beginning we present two simple equations, which are a direct consequence of the equations (\ref{q-costr-x1-c}), (\ref{q-costr-z1-c}).
\begin{description}
\item[1)] Contraction (\ref{q-costr-x1-c}) (or (\ref{q-costr-z1-c})) with $\partial^{\alpha}$ (or
$\bar\partial^{\dot\alpha}$) and use of
two-spinor identities $\partial^{\alpha}\partial_{\alpha}{=}\bar\partial^{\dot\alpha}\bar\partial_{\dot\alpha}{=}0$ leads to the relation
\begin{equation}\label{Tr-Psi-a}
\partial^{\alpha\dot\alpha}\partial_{\alpha}\bar\partial_{\dot\alpha}\,\Phi_0=0\,.
\end{equation}
The equation (\ref{Tr-Psi-a}) provides the generalized Lorentz conditions for the component fields defined by (\ref{WF-0-a})
\begin{equation}\label{Tr-comp-a}
\partial^{\alpha_1\dot\beta_1}\,\phi^{(k,n)}_{\alpha_1...\alpha_k\,\dot\beta_1...\dot\beta_n}(x)=0\,,\qquad k,n\geq 1\,.
\end{equation}
For example, if $n=k=1$ the equation (\ref{Tr-comp-a})
is the standard Lorentz condition for the four-vector field.

\item[2)] By considering the difference of the equations (\ref{q-costr-x1-c}) multiplied by $\bar\partial_{\dot\alpha}$
and the equations  (\ref{q-costr-z1-c}) multiplied by $\partial_{\alpha}$ we obtain
\begin{equation}\label{Eq-Psi-a}
\partial_{\dot\alpha}^{\beta}\,\partial_{\alpha}\partial_{\beta}\,\Phi_0=
\partial^{\dot\beta}_{\alpha}\,\bar\partial_{\dot\alpha}\bar\partial_{\dot\beta}\,\Phi_0\,.
\end{equation}
From (\ref{Eq-Psi-a}) we get the following equations for the component fields
\begin{equation}\label{Eq-comp-a}
\partial_{\dot\beta_{n-1}}^{\,\alpha_k}\,\phi^{(k,n-2)}_{\alpha_1...\alpha_{k-1}\alpha_k\,\dot\beta_1...\dot\beta_{n-2}}=
\partial^{\,\dot\beta_n}_{\alpha_{k-1}}\,\phi^{(k-2,n)}_{\alpha_1...\alpha_{k-2}\,\dot\beta_1...\dot\beta_{n-1}\dot\beta_n} \,,\qquad k,n\geq 2\,.
\end{equation}
The equations for lowest component fields described by Lorentz spins $(2,0)+(0,2)$ are as follows
\begin{equation}\label{Eq-comp-M-a}
\partial_{\dot\alpha}^{\,\beta}\,\phi^{(2,0)}_{\alpha\beta}=
\partial^{\,\dot\beta}_{\alpha}\,\phi^{(0,2)}_{\dot\alpha\dot\beta}\,.
\end{equation}
They represent half of the Maxwell equations (self-dual part) for the free electromagnetic field strength.
\end{description}

Full set of the equations for component fields of the HS master field (\ref{WF-0-a}), generated by the conditions
(\ref{q-costr-x1-a}), (\ref{q-costr-z1-a}), are
\begin{equation}\label{Eq-Dir-comp1-a}
\partial_{\dot\beta_{n+1}}^{\,\alpha_k}\,\phi^{(k,n)}_{\alpha_1...\alpha_k\,\dot\beta_1...\dot\beta_{n}}=
iem\,x^{\,\alpha_k\dot\beta_{n+2}}\,\phi^{(k,n+2)}_{\alpha_1...\alpha_{k}\,\dot\beta_1...\dot\beta_{n+1}\dot\beta_{n+2}} \,,\qquad k\geq 1\,;
\end{equation}
\begin{equation}\label{Eq-Dir-comp2-a}
\partial^{\,\dot\beta_n}_{\alpha_{k+1}}\,\phi^{(k,n)}_{\alpha_1...\alpha_{k}\,\dot\beta_1...\dot\beta_n}=
iem\,x^{\alpha_{k+2}\dot\beta_{n}}\,\phi^{(k+2,n)}_{\alpha_1...\alpha_{k+1}\alpha_{k+2}\,\dot\beta_1...\dot\beta_{n}} \,,\qquad n\geq 1\,.
\end{equation}
They represent the Maxwell-invariant generalizations of well-known Dirac-Pauli-Fierz equations~\cite{Dirac,PauliFierz}.
\vspace{0.01cm}

{}For completing the analysis of constraints it is necessary to impose on the wave function the scalar constraint
(\ref{constr-x3}). This additional condition leads to additional equation for first scalar component of the
wave function (\ref{WF-0-a}). The constraint (\ref{constr-x3}) implies the following equation (see details in~\cite{FedLuk})
\begin{equation}\label{q-constr-x3-Psi-a}
\left[\partial_{\alpha\dot\alpha}\partial^{\alpha\dot\alpha}
-2e^2m^2\left(\partial_{\alpha}x^{\alpha\dot\alpha}\bar\partial_{\dot\alpha}\right)^2\,\right]\Phi_0= 0
\end{equation}
for the HS master field  (\ref{WF-0-a}).
The equation
(\ref{q-constr-x3-Psi-a}) provides the following infinite set of field equations for the HS component fields
\begin{equation}\label{Eq-KG-comp-a}
\partial^{\gamma\dot\gamma}\partial_{\gamma\dot\gamma}\,\phi^{(k,n)}_{\alpha_1...\alpha_{k}\,\dot\beta_1...\dot\beta_n}=
2e^2m^2\,x^{\alpha_{k+1}\dot\beta_{n+1}}\,x^{\alpha_{k+2}\dot\beta_{n+2}}\,
\phi^{(k+2,n+2)}_{\alpha_1...\alpha_{k}\alpha_{k+1}\alpha_{k+2}\,\dot\beta_1...\dot\beta_{n}\dot\beta_{n+1}\dot\beta_{n+2}} \,,
\end{equation}
which are
the Maxwell-invariant generalization of massless Klein-Gordon equation. One can show that the equations (\ref{Eq-KG-comp-a}) for
all component fields $\phi^{(k,n)}(x)$, except $\phi^{(0,0)}(x)$, are the consequences of the Maxwell-Dirac equations (\ref{Eq-Dir-comp1-a}), (\ref{Eq-Dir-comp2-a}).
The only equation in  (\ref{Eq-KG-comp-a}) which is not following from (\ref{Eq-Dir-comp1-a}), (\ref{Eq-Dir-comp2-a}),
describe the Maxwell-Klein-Gordon equation
\begin{equation}\label{Eq-KG-comp-0-a} \partial^{\gamma\dot\gamma}\partial_{\gamma\dot\gamma}\,\phi^{(0,0)}=
2e^2m^2\,x^{\alpha_{1}\dot\beta_{1}}\,x^{\alpha_{2}\dot\beta_{2}}\, \phi^{(2,2)}_{\alpha_1\alpha_{2}\,\dot\beta_1\dot\beta_{2}}
\end{equation}
for the scalar field $\phi^{(0,0)}(x)$.
\vspace{0.3cm}

The Maxwell-Weyl equations (\ref{Eq-Dir-comp1-a}), (\ref{Eq-Dir-comp2-a})
and the Maxwell-Klein-Gordon equations (\ref{Eq-KG-comp-0-a} link
the fields $\phi^{(k+2p,n+2r)}(x)$  with fixed $k$, $n$ and
$p=0,1,2,...$, $r=0,1,2,...$. We obtain infinite-dimensional Maxwell-HS multiplets with minimal Lorentz spin
$(\frac{k}{2},\frac{n}{2})$ described by the field $\phi^{(k,n)}(x)$ (see the Maxwell symmetry transformations of the components
of Maxwell-HS multiplets in~\cite{FedLuk}).
In particular, we get the maximal (with largest number of components) bosonic Maxwell HS multiplet with scalar field $\phi^{(0,0)}(x)$ and the
component fields with even Lorents spins $(j_1,j_2)=(p,r)$. Further, there are two maximal fermionic multiplets: chiral, with Maxwell-Weyl field
$\phi^{(1,0)}_{\alpha}(x)$, and the antichiral one, with Maxwell-Weyl field $\phi^{(0,1)}_{\dot\alpha}(x)$.

In limit $e\to 0 $ the equations (\ref{q-costr-x1-a}), (\ref{q-costr-z1-a}) coincide with massless Dirac-Pauli-Fierz equations for standard massless conformal HS
fields
\begin{equation}\label{Dir-m0-a}
e=0\,:\qquad\qquad
\partial^{\alpha_1\dot\gamma}\,\phi^{(k,n)}_{\alpha_1...\alpha_k\,\dot\beta_1...\dot\beta_n}(x)=0\,,\qquad
\partial^{\gamma\dot\beta_1}\,\phi^{(k,n)}_{\alpha_1...\alpha_k\,\dot\beta_1...\dot\beta_n}(x)=0\,.
\end{equation}
Moreover, in the limit $e\to 0 $ all the component fields, including `mixed' fields
$\phi^{(k>0,n>0)}_{\alpha_1...\alpha_k\,\dot\beta_1...\dot\beta_n}(x)$, are independent.
This property differs from the case of standard HS particle \cite{BL,BLS,Vas} where the spectrum of independent fields contains only the scalar field $\phi^{(0,0)}(x)$,
the spin-tensor fields with undotted indices $\phi^{(k,0)}_{\alpha_1...\alpha_k}(x)$ ($k>0$) and spin-tensor field
with dotted indices $\phi^{(0,n)}_{\dot\alpha_1...\dot\alpha_n}(x)$ ($n>0$).
In standard HS particle model due to the presence of one more first class constraint the `mixed' fields
$\phi^{(k>0,n>0)}_{\alpha_1...\alpha_k\,\dot\beta_1...\dot\beta_n}(x)$ are expressed by space-time
derivatives of the fields $\phi^{(0,0)}(x)$, $\phi^{(k>0,0)}_{\alpha_1...\alpha_k}(x)$, $\phi^{(0,n>0)}_{\dot\alpha_1...\dot\alpha_n}(x)$
which follow from four components of unfolded equation
$\left( \partial_{\alpha\dot\beta}+ i  \partial_{\alpha}\bar\partial_{\dot\beta}\right)\Phi=0$.
In our case the ``mixed'' fields stay independent because there are only three linearly independent components of unfolded equation
relating Maxwell-HS fields with different Lorentz spins $(k,n)$.

\setcounter{equation}{0}
\section{HS field equations in dual representation}

\quad\, In this section we consider the representation with the generalized coordinates
$x^{\alpha\dot\beta}$, $f_{\alpha\beta}$, $\bar f_{\dot\alpha\dot\beta}$, $y^\alpha,\bar y^{\dot\alpha}$,
the momenta (\ref{op-real1-a}) and
\begin{equation}\label{op-real1}
z^{\alpha\beta}=i\frac{\partial\,\,}{\partial f_{\alpha\beta}}\,,\qquad \bar z^{\dot\alpha\dot\beta}=i\frac{\partial\,\,}{\partial \bar
f_{\dot\alpha\dot\beta}}\,.
\end{equation}
The wave function has the following functional form
\begin{equation}\label{WF}
\tilde\Psi =\tilde\Psi(x^{\alpha\dot\beta},f_{\alpha\beta}, \bar f_{\dot\alpha\dot\beta}, y^\alpha,\bar y^{\dot\alpha})
\end{equation}
and is related with the formula (\ref{WF-a}) by Fourier transform in the tensorial variables $(f_{\alpha\beta}, \bar f_{\dot\alpha\dot\beta})$.

First class constraints  (\ref{costr-x1})-(\ref{costr-y1}) yield in this representation the following equations
\begin{eqnarray}\label{q-costr-x1}
i D_{\alpha\dot\beta}\bar\partial^{\dot\beta}\,\tilde\Psi&=&\left(\partial_{\alpha\dot\beta}+ i
\,e\,f_{\alpha\gamma}\,x_{\dot\beta}^{\gamma}\right)\bar\partial^{\dot\beta}\,\tilde\Psi= 0\, ,\\[5pt] \label{q-costr-x1a} i
D_{\beta\dot\alpha}\partial^{\beta}\,\tilde\Psi&=&\left(\partial_{\beta\dot\alpha}+ i \,e\,\bar f_{\dot\alpha\dot\gamma} \,x_{\beta}^{\dot\gamma}
\right)\partial^{\beta}\, \tilde\Psi= 0\, ,
\end{eqnarray}
\begin{equation}\label{q-costr-y1}
\Big(f_{\alpha\beta}+
 m\partial_{\alpha} \partial_{\beta}\Big)\tilde\Psi=0\,,\qquad\qquad
\Big(\bar f_{\dot\alpha\dot\beta}+ m \bar\partial_{\dot\alpha}\bar\partial_{\dot\beta}\Big)\tilde\Psi=0\,.
\end{equation}

{}For the analysis of equations (\ref{q-costr-x1})-(\ref{q-costr-y1}) we introduce Taylor expansion of
the wave function (\ref{WF}) with respect to spinor variables $y^{\alpha}$ and $\bar y^{\dot\alpha}$
\begin{equation}\label{WF-1}
\tilde\Psi(x,f, \bar f, y,\bar y)= \sum_{k,n=0}^{\infty}\frac{1}{k!\,n!}\,y^{\alpha_1}...y^{\alpha_k}\, \bar
y^{\dot\beta_1}...\bar y^{\dot\beta_n} \,\tilde\Psi^{(k,n)}_{\alpha_1...\alpha_k\,\dot\beta_1...\dot\beta_n}(x,f,\bar f)\,.
\end{equation}
Then the
constraints (\ref{q-costr-y1}) lead to expression of all components $\tilde\Psi^{(k,n)}$, $k>1$, $n>1$ in terms of
four HS master fields $\tilde\Psi^{(0,0)}$, $\tilde\Psi^{(1,0)}$, $\tilde\Psi^{(0,1)}$ and $\tilde\Psi^{(1,1)}$.
We obtain the equations
$$
\tilde\Psi^{(2,0)}_{\alpha\beta}= -
m^{-1}f_{\alpha\beta} \tilde\Psi^{(0,0)}\,,\qquad \tilde\Psi^{(0,2)}_{\dot\alpha\dot\beta}= - m^{-1}\bar f_{\dot\alpha\dot\beta} \tilde\Psi^{(0,0)}\,,
$$
$$
\tilde\Psi^{(2,1)}_{\alpha\beta\dot\gamma}= - m^{-1}f_{\alpha\beta} \tilde\Phi^{(0,1)}_{\dot\gamma}\,,\qquad
\tilde\Psi^{(1,2)}_{\gamma\dot\alpha\dot\beta}= - m^{-1}\bar f_{\dot\alpha\dot\beta} \tilde\Psi^{(1,0)}_{\gamma}\,, $$ $$ \qquad \qquad
\tilde\Psi^{(3,1)}_{\alpha\beta\gamma\dot\gamma}= - m^{-1}f_{\alpha\beta} \tilde\Psi^{(1,1)}_{\gamma\dot\gamma}\,,\qquad
\tilde\Psi^{(1,3)}_{\gamma\dot\alpha\dot\beta\dot\gamma}= - m^{-1}\bar f_{\dot\alpha\dot\beta} \tilde\Psi^{(1,1)}_{\gamma\dot\gamma}\,,\qquad \mbox{etc}.
$$
In the expansion  (\ref{WF-1}) the independent HS master fields are
\begin{description}
\item[i)] one ``generalized spin zero'' field
\begin{equation}\label{g-spin0}
\tilde\Psi^{(0,0)}(x,f,\bar f)\,,
\end{equation}
\item[ii)] two ``generalized spin half'' fields
\begin{equation}\label{g-spin12}
\tilde\Psi^{(1,0)}_{\alpha}(x,f,\bar f)\,,\qquad \tilde\Psi^{(0,1)}_{\dot\beta}(x,f,\bar f)
\end{equation}
\item[iii)] one ``generalized spin one'' field
\begin{equation}\label{g-spin1}
\tilde\Psi^{(1,1)}_{\alpha\dot\beta}(x,f,\bar f)\,.
\end{equation}
\end{description}
The equations (\ref{q-costr-y1}) provide for the fields (\ref{g-spin0}), (\ref{g-spin12}), (\ref{g-spin1}) the kinematic
constraints:
\begin{equation}\label{con-spin0}
f^{\alpha\beta}f_{\alpha\beta}\tilde\Psi^{(0,0)}(x,f,\bar f)=0\,,\qquad \bar f^{\dot\alpha\dot\beta}\bar f_{\dot\alpha\dot\beta}\tilde\Psi^{(0,0)}(x,f,\bar
f)=0\,,
\end{equation}
\begin{equation}\label{con-spin12}
f_{\alpha}{}^{\beta}\tilde\Psi^{(1,0)}_{\beta}(x,f,\bar f)=0\,,\qquad \bar
f_{\dot\alpha}{}^{\dot\beta}\tilde\Psi^{(0,1)}_{\dot\beta}(x,f,\bar f)=0\,,
\end{equation}
\begin{equation}\label{con-spin1}
f_{\alpha}{}^{\beta}\tilde\Psi^{(1,1)}_{\beta\dot\beta}(x,f,\bar f)=0\,,\qquad \bar
f_{\dot\alpha}{}^{\dot\beta}\tilde\Psi^{(1,1)}_{\beta\dot\beta}(x,f,\bar f)=0\,.
\end{equation}

{}From the equations (\ref{q-costr-x1}), (\ref{q-costr-x1a}) one obtains for the unconstrained component (\ref{g-spin12}), (\ref{g-spin1}) the relations
\begin{equation}\label{Dir-12}
\left(\partial^{\alpha\dot\beta}- i
\,e\,f^{\alpha\gamma}\,x^{\dot\beta}_{\gamma}\right)\tilde\Psi^{(0,1)}_{\dot\beta}(x,f,\bar f)= 0\, ,\qquad \left(\partial^{\beta\dot\alpha}- i \,e\,\bar
f^{\dot\alpha\dot\gamma} \,x^{\beta}_{\dot\gamma} \right)\tilde\Psi^{(1,0)}_{\beta}(x,f,\bar f)= 0\, ,
\end{equation}
\begin{equation}\label{Dir-1}
\left(\partial^{\alpha\dot\beta}- i  \,e\,f^{\alpha\gamma}\,x^{\dot\beta}_{\gamma}\right)\tilde\Psi^{(1,1)}_{\beta\dot\beta}(x,f,\bar f)= 0\, ,\qquad
\left(\partial^{\beta\dot\alpha}- i \,e\,\bar f^{\dot\alpha\dot\gamma} \,x^{\beta}_{\dot\gamma} \right)\tilde\Psi^{(1,1)}_{\beta\dot\beta}(x,f,\bar f)= 0\,
,
\end{equation}
which have the form of the Dirac equations in a constant electromagnetic field, with electromagnetic potential
$\mathscr{A}_{\mu}=f_{\mu\nu}x^\nu$. As generalization of the standard approach for Dirac spin-half field,
the wave functions in (\ref{Dir-12}), (\ref{Dir-1}) depend also on
continuous electromagnetic field strength coordinates $f_{\alpha\beta}$, $\bar f_{\dot\alpha\dot\beta}$.
We do not see yet the relation of our description of
HS fields interacting with constant EM field to the approaches proposed in recent papers
on EM coupling of HS fields \cite{PRSag,BuchSZ},
however to find such a link would be desirable.

Generalized spin-zero field $\tilde\Psi^{(0,0)}(x,f,\bar f)$ is described by generalized Klein-Gordon equation, which follows from the constraint
\begin{equation}\label{KG-gen}
\hat T^{\alpha\dot\alpha}\hat T_{\alpha\dot\alpha}\,\tilde\Psi\approx 0\,.
\end{equation}
Taking into account that
\begin{equation}\label{KG-op}
\begin{array}{rcl}
\hat T^{\alpha\dot\alpha}\hat T_{\alpha\dot\alpha}&=&
-\partial^{\alpha\dot\alpha}\partial_{\alpha\dot\alpha} +2ie\left(f^{\alpha\beta}x^{\dot\alpha}_{\beta} +\bar
f^{\dot\alpha\dot\beta}x^{\alpha}_{\dot\beta}\right)\partial_{\alpha\dot\alpha}\\[6pt] && +{\textstyle\frac12}\,e^2\left(f^{\alpha\beta}f_{\alpha\beta} +
\bar f^{\dot\alpha\dot\beta}\bar f_{\dot\alpha\dot\beta}\right)x^{\gamma\dot\gamma}x_{\gamma\dot\gamma} -2e^2f_{\alpha\beta}\bar
f_{\dot\alpha\dot\beta}x^{\beta\dot\alpha}x^{\alpha\dot\beta}\\[6pt] && -2\left(i\partial^{\alpha\dot\alpha}+ef^{\alpha\beta}x^{\dot\alpha}_{\beta} + e\bar
f^{\dot\alpha\dot\beta}x^{\alpha}_{\dot\beta}\right)\partial_{\alpha}\bar\partial_{\dot\alpha}\,,
\end{array}
\end{equation}
we obtain the following generalized
Klein-Gordon equation for ``generalized spin zero'' field
\begin{equation}\label{KG-0}
\left[ -\Box +2ie\left(f^{\alpha\beta}x^{\dot\alpha}_{\beta} +\bar
f^{\dot\alpha\dot\beta}x^{\alpha}_{\dot\beta}\right)\partial_{\alpha\dot\alpha} +{\textstyle\frac12}\,e^2\Big( f^2+\bar f^2 \Big)x^{2}
-2e^2f_{\alpha\beta}\bar f_{\dot\alpha\dot\beta}x^{\beta\dot\alpha}x^{\alpha\dot\beta}\right]\tilde\Psi^{(0,0)}=0\,,
\end{equation}
where
$\Box:=\partial^{\alpha\dot\alpha}\partial_{\alpha\dot\alpha}$, $x^2:=x^{\alpha\dot\alpha}x_{\alpha\dot\alpha}$, $f^2:=f^{\alpha\beta}f_{\alpha\beta}$,
$\bar f^2:=\bar f^{\dot\alpha\dot\beta}\bar f_{\dot\alpha\dot\beta}$. It should be emphasized that due to the equations (\ref{Dir-1}) and the constraints
(\ref{con-spin1}) the last term in the operator (\ref{KG-op}) does not contribute to the equation (\ref{KG-0}) and finally we obtain standard Klein-Gordon
equation with coupling to constant EM field.

{}From the equations (\ref{Dir-12}), (\ref{Dir-1}) and (\ref{KG-0}) follows that the link of different spins is due to EM coupling
proportional to electric charge $e$. Further one can show that if the torsion
in six tensorial dimensions of Maxwell space-time depends only on the $D=4$ space-time
coordinates it can be interpreted as a coupling to $D=4$ Abelian gauge potential. Let us introduce the following ``block-diagonal'' 10-bein
$E_{AB}{}^{CD}=\left( \delta_\alpha^\gamma \delta_{\dot\beta}^{\dot\delta}, E_{\alpha\dot\beta}{}^{\gamma\delta},
E_{\alpha\dot\beta}{}^{\dot\gamma\dot\delta}\right)$ in the tensorial space $\left( x^{\alpha\dot\beta}, z^{\alpha\beta}, \bar
z^{\dot\alpha\dot\beta}\right)$ and corresponding covariant derivatives
\begin{equation}\label{cov-1-N}
\nabla_{\alpha\dot\beta}=\partial_{\alpha\dot\beta} +
E_{\alpha\dot\beta}{}^{\gamma\delta}(x)\nabla_{\gamma\delta}+ E_{\alpha\dot\beta}{}^{\dot\gamma\dot\delta}(x)\nabla_{\dot\gamma\dot\delta}\,,
\end{equation} \begin{equation}\label{cov-2-N} \nabla_{\alpha\beta}=\partial_{\alpha\beta}\,,\qquad
\nabla_{\dot\alpha\dot\beta}=\partial_{\dot\alpha\dot\beta}\,.
\end{equation}
If we consider the plane wave solutions in additional dimensions, one can
replace (see (\ref{op-real1-a})) the derivatives  (\ref{cov-2-N}) by constant tensors $f_{\alpha\beta}$, $\bar f_{\dot\alpha\dot\beta}$ which represent
additional tensorial momenta. In such a case the derivative  (\ref{cov-1-N}) can be written as the Abelian gauge-covariant derivative
\begin{equation}\label{cov-3-N}
\nabla_{\alpha\dot\beta}=\partial_{\alpha\dot\beta} + e\mathscr{A}_{\alpha\dot\beta}(x)\,,
\end{equation}
where
$e\mathscr{A}_{\alpha\dot\beta}(x)=E_{\alpha\dot\beta}{}^{\gamma\delta}(x)f_{\gamma\delta}+ E_{\alpha\dot\beta}{}^{\dot\gamma\dot\delta}(x)\bar
f_{\dot\gamma\dot\delta}$. In Maxwell tensorial space additional tensorial coordinates are twisted by a constant torsion
proportional to $e$, the functions
$E_{\alpha\dot\beta}{}^{\gamma\delta}(x)$ and $E_{\alpha\dot\beta}{}^{\dot\gamma\dot\delta}(x)$ are linear in $x$, and we obtain in (\ref{cov-3-N}) the
Abelian gauge field four-potential $\mathscr{A}_{\alpha\dot\beta}$ describing constant electromagnetic field strength ($\bar
f_{\dot\alpha\dot\beta}=(f_{\alpha\beta})^\dagger$)
\begin{equation}\label{const-A}
\mathscr{A}_{\alpha\dot\beta}=
f_{\alpha}^{\,\gamma}x_{\gamma\dot\beta}+\bar f_{\dot\beta}^{\,\dot\gamma}x_{\alpha\dot\gamma}\,.
\end{equation}
It can be added that the space-time translations
$x_{\alpha\dot\beta}\to x_{\alpha\dot\beta} +a_{\alpha\dot\beta}$ modify (\ref{const-A}) by a constant term, which can be however compensated by the
standard Abelian gauge transformation of $\mathscr{A}_{\alpha\dot\beta}$, what leads finally to the translational invariance of the covariant derivative
(\ref{cov-3-N}).

The solutions of the equations  (\ref{con-spin0})-(\ref{con-spin1}) can be represented as the Fourier transforms
\begin{eqnarray}
\tilde\Psi^{(0,0)}(x,f,\bar f)&=&\int d{}^{\,6}\!z\, e^{-i(fz+\bar f \bar z)} \,\Psi^{(0,0)}(x,z,\bar z) \,,\label{ser0}\\
\tilde\Psi^{(1,0)}_{\gamma}(x,f,\bar f)&=&\int d{}^{\,6}\!z\, e^{-i(fz+\bar f \bar z)} \,\Psi^{(1,0)}_{\gamma}(x,z,\bar z) \,,\label{ser12a}\\
\tilde\Psi^{(0,1)}_{\dot\gamma}(x,f,\bar f)&=&\int d{}^{\,6}\!z\, e^{-i(fz+\bar f \bar z)} \,\Psi^{(0,1)}_{\dot\gamma}(x,z,\bar z)\,,\label{ser12b}\\
\tilde\Psi^{(1,1)}_{\gamma\dot\gamma}(x,f,\bar f)&=&\int d{}^{\,6}\!z\, e^{-i(fz+\bar f \bar z)} \,\Psi^{(1,1)}_{\gamma\dot\gamma}(x,z,\bar z)
\,,\label{ser1}
\end{eqnarray}
where
\begin{eqnarray}
\Psi^{(0,0)}(x,z,\bar z)&=&
\sum_{k,n=0}^{\infty}\frac{1}{k!n!}\,z^{\alpha_1\beta_1}...z^{\alpha_k\beta_k}\, \bar z^{\dot\alpha_1\dot\beta_1}...\bar z^{\dot\alpha_n\dot\beta_n}
\,\phi^{(2k,2n)}_{(\alpha_1\beta_1...\alpha_k\beta_k)\,(\dot\alpha_1\dot\beta_1...\dot\alpha_n\dot\beta_n)}(x)\,,\label{com-ser0}\\
\Psi^{(1,0)}_{\gamma}(x,z,\bar z)&=& \sum_{k,n=0}^{\infty}\frac{1}{k!n!}\,z^{\alpha_1\beta_1}...z^{\alpha_k\beta_k}\, \bar
z^{\dot\alpha_1\dot\beta_1}...\bar z^{\dot\alpha_n\dot\beta_n}
\,\phi^{(2k+1,2n)}_{(\gamma\alpha_1\beta_1...\alpha_k\beta_k)\,(\dot\alpha_1\dot\beta_1...\dot\alpha_n\dot\beta_n)}(x)\,,\label{com-ser12a}\\
\Psi^{(0,1)}_{\dot\gamma}(x,z,\bar z)&=& \sum_{k,n=0}^{\infty}\frac{1}{k!n!}\,z^{\alpha_1\beta_1}...z^{\alpha_k\beta_k}\, \bar
z^{\dot\alpha_1\dot\beta_1}...\bar z^{\dot\alpha_n\dot\beta_n}
\,\phi^{(2k,2n+1)}_{(\alpha_1\beta_1...\alpha_k\beta_k)\,(\dot\gamma\dot\alpha_1\dot\beta_1...\dot\alpha_n\dot\beta_n)}(x)\,,\label{com-ser12b}\\
\Psi^{(1,1)}_{\gamma\dot\gamma}(x,z,\bar z)&=& \sum_{k,n=0}^{\infty}\frac{1}{k!n!}\,z^{\alpha_1\beta_1}...z^{\alpha_k\beta_k}\, \bar
z^{\dot\alpha_1\dot\beta_1}...\bar z^{\dot\alpha_n\dot\beta_n}
\,\phi^{(2k+1,2n+1)}_{(\gamma\alpha_1\beta_1...\alpha_k\beta_k)\,(\dot\gamma\dot\alpha_1\dot\beta_1...\dot\alpha_n\dot\beta_n)}(x)\label{com-ser1}
\end{eqnarray}
have polynomial dependence on $z$, $\bar z$ and the component HS fields $\phi^{(k,n)}_{(\alpha...)\,(\dot\alpha...)}$ are symmetric with
respect to all undotted and dotted spinor indices (compare with HS fields in (\ref{WF-0-a})
considered in previous subsection). The fields (\ref{ser0})-(\ref{ser1}) have
complicated, nonpolynomial dependence on the variables $f$, $\bar f$, with generalized Dirac equations (see (\ref{Dir-12})-(\ref{Dir-1})) and
Klein-Gordon equation (\ref{KG-0}) providing the explicit functional dependence of the Maxwell-HS component fields on the constant
values of EM field.

\setcounter{equation}{0} \section{Outlook}

\quad\, In this paper we considered the spinorial particle model in ten-dimensional
tensorial space-time with torsion described by the tensorial coset space $X
=(x^{\alpha\dot\alpha},z^{\alpha\beta}, \bar z^{\dot\alpha\dot\beta} )$ of $D=4$ Maxwell group and
additional spinorial variables $\lambda^{\alpha}$,
$\bar\lambda^{\dot\alpha}$.
We performed the canonical quantization of the model with supplemented kinetic term for $\lambda^{\alpha}$,
$\bar\lambda^{\dot\alpha}$. By using the phase space formulation we specified the set of first and second class constraints. It appears that in first quantized
theory the first class constraints will describe the set of field equations for new higher spin multiplets in the tensorial space $X$
defining new HS Maxwell dynamics. Such equations describe the
generalization of the known ``unfolded equations'' \cite{BLS,Vas,Vas1} for massless HS free fields with flat space-time derivatives
$\partial_{\alpha\dot\beta}$ replaced by the Maxwell-covariant derivatives $D_{\alpha\dot\beta}$ (see (\ref{M-der})).
Note that the Maxwell-covariant
description of $D=4$ Maxwell-HS fields requires the presence of particular space-time-dependent coupling terms between different spin fields which can be also
interpreted as following from the electromagnetic covariantization of space-time derivatives in the presence of
constant EM background field strength.

As an interesting question which we plan to study is the derivation of massive HS fields from twistorial model of Shirafuji type, with
two spinors which are necessary in $D=4$ in order to define in spinorial framework the time-like four-momentum.
The idea of describing massive particles by multispinors is well-known from the consideration of Penrose and his collaborators
\cite{Penr1,PenrM,Penr2,Hugh,Bette} and further was considered as well as in the supersymmetric case in \cite{SV,FZ,FedZ1,FedZ2,BALM,AFLM,FFLM}.
In order to illustrate the derivation of massive spinorial model we shall argue following  \cite{SV}
that $D=3$ massive Shirafuji model can be obtained by dimensional reduction of $D=4$ spinorial model considered in  \cite{FedLuk}.

For that purpose we should perform the decompositions of $D=4$ tensors in term of corresponding $D=3$ objects  \cite{SV}.
$D=4$ four-vector $X_{\alpha\dot\beta}$ has the following representation
\begin{equation}\label{x}
X_{\alpha\dot\beta}=x_{\alpha\beta}-ix_{2}\epsilon_{\alpha\beta}\,,\qquad X^{\dot\alpha\beta}=x^{\alpha\beta}-ix_{2}\epsilon^{\alpha\beta}\,,
\end{equation}
where the tensor $x^{\alpha\beta}=x^{\beta\alpha}=(\overline{x^{\alpha\beta}})$ describe $D=3$ three-vector and $\epsilon_{\alpha\beta}=-\epsilon_{\beta\alpha}$,
$\epsilon^{\alpha\beta}=-\epsilon^{\beta\alpha}$, $\epsilon_{12}=\epsilon^{21}=1$ represent $D=3$ invariant tensors
raising or lowering spinor indices: $x_{\alpha}^{\gamma}=\epsilon_{\alpha\beta}x^{\beta\gamma}$.
The $D=3$ decomposition of $z_{\mu\nu}$ after using the spinorial notation looks as follows:
\begin{equation}\label{z}
z^{\alpha\beta}=z_1^{\alpha\beta}+iz_2^{\alpha\beta}\,,\qquad \bar z^{\dot\alpha\dot\beta}=z_1^{\alpha\beta}-iz_2^{\alpha\beta}\,,
\end{equation}
where $(\overline{z_1^{\alpha\beta}})=z_1^{\alpha\beta}$, $(\overline{z_2^{\alpha\beta}})=z_2^{\alpha\beta}$ are real.
Further $D=4$ Weyl spinor
\begin{equation}\label{l}
\lambda^{\alpha}=u^{\alpha}+iv^{\alpha}\,,\qquad \bar\lambda^{\dot\alpha}=u^{\alpha}-iv^{\alpha}
\end{equation}
provides two $D=3$ Majorana (real) spinors $(\overline{u^{\alpha}})=u^{\alpha}$, $(\overline{v^{\alpha}})=v^{\alpha}$. Similarly we have
\begin{equation}\label{y}
y^{\alpha}={\textstyle \frac12}\,(\rho^{\alpha}-i\varrho^{\alpha})\,,\qquad
\bar y^{\dot\alpha}={\textstyle \frac12}\,(\rho^{\alpha}+i\varrho^{\alpha})\,,
\end{equation}
where $(\overline{\rho^{\alpha}})=\rho^{\alpha}$, $(\overline{\varrho^{\alpha}})=\varrho^{\alpha}$ are Majorana spinors.

Let us consider the $D=4$ Shirajuji action \cite{Shir} for spinless massless particle
\begin{equation}\label{4-3}
\begin{array}{rcl}
S^{D=4}_{m=0}&=&{\displaystyle\int} d\tau \lambda_{\alpha}\bar\lambda_{\dot\beta}\dot X^{\dot\beta\alpha}
\\[6pt]
&=& {\displaystyle\int} d\tau\Big[ \left(u_{\alpha}u_{\beta}+v_{\alpha}v_{\beta} \right)\,\dot x^{\alpha\beta} + 2( u_{\alpha}v^{\alpha})\,\dot \varphi\Big]\,,
\end{array}
\end{equation}
where we denote $\varphi\equiv x_2$ (see  (\ref{x})). From (\ref{4-3}) follows by varying of $\varphi$ that
$\frac{d}{d\tau}( u_{\alpha}v^{\alpha})=0$. We will further fix the constant on-shell value of $( u_{\alpha}v^{\alpha})$
by the replacement in (\ref{4-3})
\begin{equation}\label{repl}
2( u_{\alpha}v^{\alpha})\,\dot \varphi\qquad\rightarrow\qquad -\lambda ( u_{\alpha}v^{\alpha}-m)\,.
\end{equation}
Because the three-momentum in (\ref{4-3}) is given by the formula
\begin{equation}\label{3mom}
p_{\alpha\beta}=u_{\alpha}u_{\beta}+v_{\alpha}v_{\beta}
\end{equation}
we get from (\ref{repl}) the mass-shell condition
\begin{equation}\label{3mass}
p^2\equiv p^\mu p_\mu ={\textstyle\frac12}\,p_{\alpha\beta}p^{\alpha\beta}=(u_{\alpha}v^{\alpha})^2=m^2\,.
\end{equation}
In such a way we obtain the spinorial description of $D=3$ massive particle model with linear mass-shell condition
\begin{equation}\label{3mass-act}
S^{D=3}_{m\neq0}=\int d\tau\Big[  p_{\alpha\beta}\dot x^{\alpha\beta} -\lambda\left( \sqrt{p^2}-m\right)\Big]\,.
\end{equation}

The model (\ref{3mass-act}) is equivalent to the standard massive particle model.
Indeed, the action (\ref{3mass-act})  after substituting the algebraic equations obtained by consecutive varying of $p_\mu$ and $\lambda$
\begin{equation}\label{eqs-ep}
p_{\mu}=2m\dot x_{\mu}/\lambda\,,\qquad \lambda= 2\sqrt{\dot x^2}
\end{equation}
provides (up to scale renormalization of mass parameter) the standard action for massive
spinless particle
\begin{equation}\label{3mass-act-sqrt}
S^{D=3}_{m\neq0}=m\int d\tau\,\sqrt{\dot x^2}\,.
\end{equation}

Following \cite{FedZ1} we write $D=3$ massive Shirafuji model as follows
\begin{equation}\label{3mass-act-Sc}
S^{D=3}_{m\neq0}= {\displaystyle\int} d\tau\Big[ \left(u_{\alpha}u_{\beta}+v_{\alpha}v_{\beta} \right)\,\dot x^{\alpha\beta}
-\lambda ( u_{\alpha}v^{\alpha}-m)\Big]
\end{equation}
with canonical three-momentum ${\displaystyle p_{\alpha\beta}=\frac{\partial L}{\partial\dot x^{\alpha\beta}}}$ given by the formula (\ref{3mom}).
One gets the field equations
\begin{equation}
\dot p_{\alpha\beta} =0\,,
\end{equation}
\begin{equation}\label{eq-3}
\dot x^{\alpha\beta}u_\beta-\lambda v^\alpha=0\,, \qquad
\dot x^{\alpha\beta}v_\beta+\lambda u^\alpha=0\,,
\end{equation}
\begin{equation}\label{mass-s}
u_{\alpha}v^{\alpha}=m\,.
\end{equation}
Assuming general formula
\begin{equation}
\dot x^{\alpha\beta}=au^\alpha u^\beta+b(u^\alpha v^\beta+v^\alpha u^\beta)+cv^\alpha v^\beta
\end{equation}
and substituting in (\ref{eq-3}) one gets after using (\ref{mass-s}) that $b=0$, $a=c=\lambda /m$ and we obtain
(see also~\cite{Bette})
\begin{equation}
\dot x^{\alpha\beta}=\lambda p^{\alpha\beta}/m\,.
\end{equation}
If we introduce $D=3$ twistors describing the following pair of $O(3,2)\cong Sp(4)$ real spinors
\begin{equation}
t_A^1=\left(
               \begin{array}{c}
                 u_\alpha \\
                 \omega^\beta \\
               \end{array}
             \right)
,\qquad
t_A^2=\left(
               \begin{array}{c}
                 v_\alpha \\
                 \mu^\beta \\
               \end{array}
             \right)
,
\end{equation}
where
\begin{equation}
\omega^\alpha=x^{\alpha\beta}u_\alpha\,,\qquad \mu^\alpha=x^{\alpha\beta}v_\alpha\,,
\end{equation}
the action (\ref{3mass-act-Sc}) can be rewritten modulo total derivative as a free bitwistorial particle model
\begin{equation}\label{3mass-act-tw}
S^{D=3}_{m\neq0}= {\displaystyle\int} d\tau\Big[ \left(u_{\alpha}\dot\omega^{\alpha}-\dot u_{\alpha}\omega^{\alpha} \right)+
\left(v_{\alpha}\dot\mu^{\alpha}-\dot v_{\alpha}\mu^{\alpha} \right)
-\lambda \,( u_{\alpha}v^{\alpha}-m)\Big]
\end{equation}
where mass-shell term break $O(3,2)$ symmetry representing $D=3$ conformal invariance by the so-called
``infinity twistor'' $I^{AB}$ \cite{Penr1,PenrM}
\begin{equation}
u_{\alpha}v^{\alpha}=t_A^1 I^{AB} t^2_B\,.
\end{equation}

In order to obtain massive $D=3$ Maxwell HS multiplets we should generalize the model (\ref{3mass-act-Sc}) in three steps:
\begin{description}
\item[i)] following \cite{FedLuk} to enlarge $D=3$ space-time to the generalized space-time with one or two additional vector variables
(see (\ref{z})) describing $D=3$ tensorial central charge coordinates;
\item[ii)] after the replacement (\ref{PP-Z}) by using Maxwell-covariant CM one-forms 
add the terms depending on constant EM field in order to obtain $D=3$ Maxwell-invariant Lagrangian;
\item[iii)] add the free action for additional spinorial
variables $\rho^\alpha$, $\varrho^\alpha$ (see (\ref{y})) in order to simplify the quantization procedure.
\end{description}

We plan to consider in our next publication the massive extension of the model
considered in \cite{FedLuk}. The construction of massive Shirafuji model one can obtain for
$D=3$ (real Majorana spinors), $D=4$ (complex Weyl spinors) or $D=6$
(quaternionic Weyl spinors), by using respective complex and quaternionic generalization of the formulae (\ref{x})-(\ref{y})
(see also \cite{KuTo,BenC}).

\section*{Acknowledgements}

\noindent We acknowledge a support from the
grant of the Bogoliubov-Infeld Programme and RFBR grants 12-02-00517, 13-02-90430 (S.F.), as
well as Polish National Center of Science (NCN) research project
No.~2011/01/ST2/03354 (J.L.). One of the author (J.L.) would like to
thank Cestmir Burdik for warm hospitality at the conference in Prague and
CERN Theory Division, where present paper has been completed.

\end{document}